\documentclass[conference]{IEEEtran}

\ifCLASSINFOpdf
  \usepackage[pdftex]{graphicx}
  \graphicspath{{./img/}}
  \DeclareGraphicsExtensions{.pdf,.jpeg,.png}
\else
\fi

\usepackage{amsmath}
\usepackage{braket}

\usepackage{url}
\usepackage{cleveref}
\usepackage{mathrsfs}

\usepackage{xcolor}

\makeatletter
\newcommand{\linebreakand}{%
  \end{@IEEEauthorhalign}
  \hfill\mbox{}\par
  \mbox{}\hfill\begin{@IEEEauthorhalign}
}
\makeatother

\begin{document}

\title{Qandle: Accelerating State Vector Simulation Using Gate-Matrix Caching and Circuit Splitting}

\author{
\IEEEauthorblockN{Gerhard Stenzel} %
\IEEEauthorblockA{\textit{Insistute for Informatics} \\
\textit{LMU Munich}\\
\textbf{gerhard.stenzel@ifi.lmu.de}}
\and
\IEEEauthorblockN{Sebastian Zielinski} %
\IEEEauthorblockA{\textit{Insistute for Informatics} \\
\textit{LMU Munich}\\
sebastian.zielinski@ifi.lmu.de}
\and
\IEEEauthorblockN{Michael Kölle} %
\IEEEauthorblockA{\textit{Insistute for Informatics} \\
\textit{LMU Munich}\\
michael.koelle@ifi.lmu.de}
\linebreakand
\IEEEauthorblockN{Philipp Altmann} %
\IEEEauthorblockA{\textit{Insistute for Informatics} \\
\textit{LMU Munich}\\
philipp.altmann@ifi.lmu.de}
\and
\IEEEauthorblockN{Jonas Nüßlein} %
\IEEEauthorblockA{\textit{Insistute for Informatics} \\
\textit{LMU Munich}\\
jonas.nuesslein@ifi.lmu.de}
\and
\IEEEauthorblockN{Thomas Gabor} %
\IEEEauthorblockA{\textit{Insistute for Informatics} \\
\textit{LMU Munich}\\
thomas.gabor@ifi.lmu.de}
}

\markboth{IEEE International Conference on Quantum Software, Summer 2024}{IEEE International Conference on Quantum Software, Summer 2024}

\maketitle

\begin{abstract}
 To address the computational complexity associated with state-vector simulation for quantum circuits, we propose a combination of advanced techniques to accelerate circuit execution. Quantum gate matrix caching reduces the overhead of repeated applications of the Kronecker product when applying a gate matrix to the state vector by storing decomposed partial matrices for each gate. Circuit splitting divides the circuit into sub-circuits with fewer gates by constructing a dependency graph, enabling parallel or sequential execution on disjoint subsets of the state vector. These techniques are implemented using the PyTorch machine learning framework. We demonstrate the performance of our approach by comparing it to other PyTorch-compatible quantum state-vector simulators. Our implementation, named \textit{Qandle}, is designed to seamlessly integrate with existing machine learning workflows, providing a user-friendly API and compatibility with the OpenQASM format. Qandle is an open-source project hosted on GitHub\footnote{https://github.com/gstenzel/qandle} and PyPI\footnote{https://pypi.org/project/qandle/}.
\end{abstract}


\IEEEpeerreviewmaketitle

\IEEEPARstart{Q}{uantum} machine learning (QML) is a rapidly expanding field that aims to combine the computational power of quantum computing with the flexibility and scalability of classical machine learning algorithms \cite{qinfo,qml1,qml2,qml3}. In recent years, machine learning has gained significant popularity and has been widely applied in various domains, including image and speech recognition, natural language processing, and recommendation systems. These applications often rely on deep learning models, which are trained on large datasets using substantial computational resources \cite{qml4,qml5,qml6,qml7,qml8,qinfo}.

Quantum machine learning seeks to harness the potential of quantum computing to solve complex optimization problems currently intractable for classical computers. By doing so, it offers a novel approach to addressing intricate challenges in machine learning and other disciplines \cite{qinfo}. However, existing quantum hardware still faces several limitations, such as hardware noise without sufficient error mitigation and correction \cite{nisq}, limited qubit connectivity \cite{torchquantum}, and a restricted number of qubits. These limitations impact the real-world performance of quantum machine learning algorithms and models.

To overcome these challenges, hybrid quantum-classical machine learning models have been developed. These models consist of classical and quantum layers, enabling training on either real hardware (with reduced noise impact due to their smaller scale) or simulators \cite{qml7}. These simulators, which run on classical hardware such as CPUs or GPUs, are used to mimic the behavior of quantum circuits. They facilitate the rapid development and training of quantum machine learning models \cite{nisq}.

As the computational complexity of quantum circuits increases exponentially with the number of qubits, the efficient performance of simulators plays a crucial role in advancing quantum machine learning. This paper introduces two novel methods, namely quantum gate matrix caching and circuit splitting, to accelerate the execution of quantum circuit simulation. We implement these methods in Qandle, a state-vector simulator we specifically designed for hybrid quantum-classical machine learning applications in conjunction with the widely adopted PyTorch library. Through a comparative analysis with existing PyTorch-compatible quantum state-vector simulators, Qandle demonstrates superior performance in terms of execution time and memory usage.

Our contributions are
\begin{enumerate}
    \item the introduction of two novel methods, namely gate matrix caching and circuit splitting, 
    \item the implementation of these methods in a new simulator and
    \item a performance comparison to existing approaches.
\end{enumerate}

This paper is structured as follows: in \cref{sec:prelims}, we introduce the required symbols and background. In \cref{sec:relatedwork}, we analyze related work and elaborate our contribution. We then present our proposed performance enhancing techniques of gate matrix caching and circuit splitting in \cref{sec:meth} and evaluate their implementation in \cref{sec:eval}. Our conclusion can be found in \cref{sec:conclusion}.

\section{Preliminaries}\label{sec:prelims}
\subsection{Symbols}
In this paper, we adopt the most significant bit first (MSb 0) notation for representing quantum states. Under this notation, the state $\ket{0000}$ corresponds to all qubits being in the state 0, while the state $\ket{0001}$ represents all qubits being in the state 0 except for the last qubit, which is in the state 1. This notation allows for a consistent and unambiguous representation of quantum states throughout our analysis. Other symbols used in this paper are summarized below.

\begin{table}[!h]
  \centering
  \begin{tabular}{|l|l|}
    \hline
    Symbol          & Description                                                  \\
    \hline
    $\ket{\varphi}$ & Quantum state of the system.                                 \\
    $S$             & State vector of $\ket{\varphi}$                              \\
    $W$             & Total number of qubits in the circuit.                       \\
    $w$             & The current qubit.                                           \\
    $R_d$           & (Matrix representation of) Gate for rotation around axis $d$ \\
    $\mathscr{R}_d$ & Matrix representation of $R_d$ on $W$ qubits.                \\
    \hline
  \end{tabular}
  \label{tab:symbols}
\end{table}

\subsection{State Vector Simulation}
The quantum state $\ket{\varphi}$ of a system with $W$ qubits can be represented as a vector of size $2^W$. This vector contains the complex probability amplitudes of each of the $2^W$ possible states, ranging from $\ket{00\dots0}$ to $\ket{11\dots1}$. Thus, it fully describes the system's state at any given time.

Quantum gates, represented by unitary matrices, are applied to the quantum state to transform it. On real quantum hardware, the state vector is not directly accessible. Instead, it can be inferred from the probabilistic measurement results of the quantum system. However, these measurements only provide an approximation of the state vector due to the inherent noisiness of the hardware in the NISQ era \cite{nisq,qinfo}.

In contrast, simulators that work with the full state vector can provide the exact state of the system at any given time. However, these simulators face a challenge when dealing with large circuits due to the exponential growth of the state vector with the number of qubits.
Due to their deterministic nature, simulators excel in building, debugging, and training variational quantum circuits.

\subsection{Hybrid Machine Learning}
In the context of quantum machine learning, hybrid machine learning refers to integrating classical and quantum machine learning algorithms. This integration can be achieved by incorporating trainable quantum circuits into larger machine learning models or by applying classical machine learning techniques to optimize quantum circuits.

Typically, quantum models in this context take the form of quantum variational circuits, which consist of several groups of gates:
\begin{enumerate}
  \item Embedding layers, which encode classical data into the quantum state of the circuit. Different embedding methods offer varying trade-offs between the expressiveness of the quantum state and the number of required qubits. Some circuit architectures employ "data re-uploading" techniques to enhance the expressiveness of the quantum state by embedding the same data points at multiple locations within the circuit, effectively reinforcing the circuit's memory of the input data.
  \item Trainable layers, which are parameterized gates whose parameters serve as the trainable weights of the quantum model. These parameters, often represented as angles of rotational gates, can be optimized using classical optimization algorithms such as gradient descent or its variants.
  \item Measurement layers, which extract relevant information encoded in the quantum state and map it to a classical output. This output can then be further processed or optimized. While simulators allow measurements at any point in the circuit, real quantum hardware typically only permits measurements as the final operation on a qubit due to its destructive nature.

\end{enumerate}
These quantum models can be treated as black boxes, enabling seamless integration into existing machine learning workflows. They can be applied to a wide range of tasks, including classification, regression, clustering, and generative modeling.

During training, the weights of the quantum models are optimized using methods such as the parameter-shift rule or classical backpropagation. The parameter-shift rule enables the calculation of the gradient of the loss function without requiring knowledge of the internal workings of the quantum circuit, making it suitable for both real quantum hardware and simulators. It approximates the gradient using the finite difference method. On the other hand, classical backpropagation, which can be efficiently deployed on state-vector simulators, treats the quantum and classical parts of the machine learning model separately and allows for different optimization algorithms and learning rates, while allowing the use of classical optimization algorithms on the quantum weights, too.

\subsection{Concept of Shapes}

The concept of shapes is employed in accordance with the notion of shape in PyTorch \cite{pytorch} tensors. A tensor is a potentially high-dimensional matrix, where the shape specifies the number of elements or sub-tensors in each dimension. For instance, a tensor with shape $(2,3,4)$ consists of two sub-matrices, each with three rows and four columns, resulting in a total of $2\cdot3\cdot4$ elements. In the context of quantum circuits, the quantum state $S$ of a system with $W$ qubits can be represented as a tensor of shape $2^W$, containing the complex probability amplitudes of each of the $2^W$ possible states $\ket{00\dots0}$ to $\ket{11\dots1}$. This can be formulated as a complex vector $S \in \mathcal{C}^{2^W}$. By employing isomorphic transformations, we can reshape the tensor to a shape of $(d_1,d_2,\dots,d_W)$, where all $d_i$ are equal to two and $W$ is the number of qubits. This changes the representation of the state from $S \in \mathcal{C}^{2^W}$ to $S \in \mathcal{C}^{2\times 2\times\dots\times2}$. Intuitively, each dimension of this tensor represents a qubit of the quantum circuit. For example, the probability amplitude of the state $\ket{010}$ is stored in the tensor at position $(0,1,0)$, which corresponds to the first element of the first dimension, the second element of the second dimension, and the first element of the third dimension.

When applying a single qubit gate, represented by a $G\in\mathcal{C}^{2\times2}$ matrix, to the $w$-th qubit, we can reshape the shape of the quantum state from $(2^W)$ to $(d_0\times d_1 \times \dots \times d_{w-1} \times d_w \times d_{w+1} \times \dots \times d_W ) $ (with all dimensions being 2), and then further rearrange the elements to $(({d_0 \times d_1 \times \dots \times d_{w-1} \times d_{w+1} \times \dots d_W }), d_w)$. This results in a tensor shape of $(2^{W-1},2)$, which can be multiplied with the gate matrix $G$ and then reshaped back to the original $S\in\mathcal{C}^{2^W}$.

In the context of machine learning, the tensor is typically extended by an additional dimension representing the batch size of the data, expanding the shape to $(B,2^W)$ or $(B,2,2,\dots,2)$ (with $B$ being the batch size, e.g., 16). This allows for processing multiple data points simultaneously during the same forward and backward passes.

\section{Related Work}\label{sec:relatedwork}
\subsection{PennyLane}
PennyLane is a Python 3 software framework for differentiable programming of quantum computers \cite{pennylane}.
It provides support for a wide range of quantum hardware and simulators,
and seamlessly integrates with machine learning libraries such as PyTorch \cite{pytorch} and Tensorflow \cite{tensorflow},
as well as other quantum software platforms including Qiskit \cite{qiskit} (see also \cref{sec:qiskit}) and Cirq \cite{cirq}.
PennyLane distinguishes between quantum nodes and classical nodes,
where quantum nodes represent the parts of the execution graph that run on a quantum device or simulator.
The framework offers an extensive collection of quantum operations, encompassing single- and multi-qubit gates, measurements, and non-unitary operations such as the Reset operation.
Furthermore, PennyLane provides built-in support for quantum chemistry simulations.

The performance of PennyLane is primarily dependent on the underlying quantum simulators,
with different backend implementations offering varying trade-offs between computational speed and supported operations.
Some simulators even support execution on NVIDIA GPUs to further enhance performance.
To expedite the execution of the same quantum circuit with different parameters, PennyLane employs caching techniques.
Additionally, most gates and simulators support batching (albeit not all), a common technique in machine learning.
PennyLane also offers circuit visualization methods and supports importing and exporting circuits in the OpenQASM 2.0 format \cite{openqasm2,pennylane}.

\subsection{Qiskit}\label{sec:qiskit}
Qiskit is a comprehensive framework for quantum computing developed by IBM \cite{qiskit,qiskittoolchain}. It offers a wide range of quantum operations, including single- and multi-qubit gates, measurements, and non-unitary operations. Qiskit provides access to real quantum hardware through the IBMQ Experience, allowing users to run their quantum circuits on IBM's quantum computers or simulators in the cloud. Local simulators are also available without the need for registration.

To optimize quantum circuits for specific quantum devices, Qiskit offers a transpiler. The transpiler adapts the circuit to hardware-specific coupling constraints, which determine the allowed combinations of qubits for CNOT gates and their directions. It also handles gate restrictions by decomposing unsupported gates into the supported set of gates for the target hardware. Additionally, gate-fusing and gate-cancellation techniques are employed to reduce the total number of gates, resulting in improved execution time and mitigating hardware noise and errors.

It is important to note that Qiskit uses the least significant bit as the first bit (LSb 0), while most other frameworks use the most significant bit as the first bit (MSb 0). This distinction can lead to confusion when using multiple frameworks simultaneously.

Qiskit's integration with the IBMQ Experience provides researchers and developers with valuable resources for exploring and experimenting with quantum computing. The combination of its extensive quantum operations, transpiler capabilities, and access to real quantum hardware makes Qiskit a powerful tool for quantum algorithm development and execution.

\subsection{TorchQuantum}
TorchQuantum \cite{torchquantum} is a recently developed framework based on PyTorch, with a focus on execution speed and parallelization. It offers seamless integration with IBM's Qiskit, allowing for easy conversion of its models to Qiskit circuits. These circuits can then be executed on real quantum hardware using IBMQ or exported to the OpenQASM format.

TorchQuantum leverages distributed GPU computing to handle large-scale circuits and batch sizes, resulting in significant performance improvements compared to PennyLane. In fact, TorchQuantum has been reported to achieve execution time improvements of up to 1000 times \cite{torchquantum}. The framework inherits the support for backpropagation and batching from the PyTorch library, enabling efficient scaling with the number of qubits and batch size.

One notable feature of TorchQuantum is its design as a tool for running QuantumNAS, a noise-adaptive search for robust quantum circuits \cite{torchquantum}. This is achieved by dividing circuits into smaller sub-circuits and optimizing them independently. The sub-circuits are then combined using an evolutionary algorithm. This approach minimizes the impact of hardware noise and therefore maximizes performance on real quantum hardware, making it highly beneficial for quantum machine learning applications.

\subsection{Contribution}
Our contribution lies in the proposal and combination of advanced techniques aimed at accelerating the execution of quantum circuits. As a result, we have developed a high-performance state-vector simulator called Qandle, which offers seamless integration into PyTorch-based machine learning workflows. Qandle demonstrates significant improvements in execution times and memory usage compared to existing frameworks such as PennyLane, Qiskit, and TorchQuantum. Notably, both of our methods are matrix-based, making them highly compatible with PyTorch's \verb|torch.compile| function, thereby further enhancing performance.

It is important to emphasize that our simulator does not aim to replace PennyLane or Qiskit. Instead, it serves as a valuable tool for quantum machine learning applications within the PyTorch ecosystem, similarly to TorchQuantum. Our simulator prioritizes efficient execution of quantum circuits on both CPU and GPU platforms, focusing on performance rather than providing advanced visualization tools or direct access to quantum hardware, unlike more mature frameworks such as PennyLane, Qiskit and TorchQuantum.

By leveraging the presented techniques gate matrix caching (\cref{sec:meth:caching}) and partial matrix decomposition (\cref{sec:meth:splitting}), our simulator optimizes the execution of gate operations on the state vector. This results in reduced computation (\cref{sec:eval:time}) and memory requirements (\cref{sec:eval:memory}) during the forward pass of the quantum circuit.

The integration of our simulator with PyTorch enables seamless incorporation of quantum circuits into machine learning models. This allows researchers and practitioners to explore the potential of quantum computing in various domains, such as quantum chemistry simulations, optimization problems, and generative modeling. Furthermore, our simulator's compatibility with the OpenQASM format facilitates interoperability with other quantum software platforms, enabling easy integration with existing quantum algorithms and libraries, thanks to its user-friendly yet powerful API (\cref{sec:meth:api}).

In summary, we combine our presented methods of gate matrix caching and circuit splitting in our presented high-performance state-vector simulator Qandle, with reduced memory usage and increased execution speed and support for just-in-time compilation, making it an attractive choice for researchers and practitioners seeking to leverage the power of quantum computing in their machine learning workflows.

\section{Performance Enhancing Techniques}\label{sec:meth}
\subsection{Gate Matrix Caching}\label{sec:meth:caching}
To improve execution times, we employ a technique we call gate matrix caching, which involves storing partial matrices of the gates. These partial matrices are decompositions of the gate matrices into two matrices with the same shape but higher sparsity. For instance, we can decompose the $R_x(\theta)$ gate into two matrices, $R_{xa}$ and $R_{xb}$, both of shape $(2,2)$, but with only two non-zero elements each.

The decomposition of the gate matrix is achieved as follows:
\begin{equation}
  \begin{split}
    R_x(\theta) & =\begin{bmatrix}\cos(\theta /2)&-i\sin(\theta /2)\\-i\sin(\theta /2)&\cos(\theta /2)\end{bmatrix} \\
    & = \begin{bmatrix}1&0\\0&1\end{bmatrix}\cdot\cos(\theta /2) + \begin{bmatrix}0&1\\1&0\end{bmatrix} \cdot -i\sin(\theta /2) \\
    & = R_{xa}\cdot\cos(\theta /2) + R_{xb}\cdot -i\sin(\theta /2)
  \end{split}
\end{equation}

The advantage of using these partial matrices is that they require fewer operations during the forward pass. Instead of allocating and filling the full gate matrix, we can simply multiply the parameters with their respective partial matrices, add the results together, and then multiply with the state vector. This reduces the computational complexity and improves the overall efficiency of the circuit.

Furthermore, the benefits of gate matrix caching are even more pronounced when working with circuits that involve multiple qubits. To expand the gate matrix $R_x$ to the full size of the state vector, we compute the Kronecker product ($\otimes$) of the partial matrices $R_{xa}$ and $R_{xb}$ with identity matrices, resulting in $\mathscr{R}_x\in\mathcal{C}^{2^W \times 2^W}$.

\begin{equation}
  \begin{split}
    \mathscr{R}_x(\theta) & = I_{2^w} \otimes R_x(\theta) \otimes I_{2^{\left(W-w\right)}} \\
    & = I_{2^w} \otimes R_{xa} \otimes I_{2^{\left(W-w\right)}} \cdot\cos(\theta /2) \\
    & + I_{2^w} \otimes R_{xb} \otimes I_{2^{\left(W-w\right)}} \cdot -i\sin(\theta /2) \\
    &= \mathscr{R}_{xa} \cdot\cos(\theta /2) + \mathscr{R}_{xb} -i\sin(\theta /2)
  \end{split}
\end{equation}

It is crucial to note the correct execution order of the Kronecker product concerning the number of states $2^w$ for the qubits before the gate and $2^{\left(W-w\right)}$ for the qubits after the gate. This order is essential for the proper reshaping of the state vector after the gate application. By utilizing the cached partial matrices $\mathscr{R}_{xa}$ and $\mathscr{R}_{xb}$, the application of the expanded matrices $\mathscr{R}_x$ is faster than computing the full gate matrix for each forward pass, which would necessitate repeated applications of the Kronecker product.

Gate matrix caching is not limited to single-qubit gates but also extends to multi-qubit gates such as CNOT and composed gate structures like the rotational gates for angle embedding layers. For these, each rotational gate is decomposed into two partial matrices. For ease of access and better hardware-level caching, the two groups of partial matrices $\mathscr{R}_a$ and $\mathscr{R}_b$ are stacked into tensors of shape $(W, 2^W, 2^W)$. During embedding, the partial embedding functions (e.g., $f_{ax}(\theta)=\cos(\theta/2)$ and $f_{bx}(\theta)=-i\sin(\theta /2)$ for the $R_x$ gate) are computed for all inputs, resulting in two vectors of shape $(W)$. These vectors are then multiplied with the partial matrices $\mathscr{R}_a$ and $\mathscr{R}_b$, respectively, along the first axis. The resulting matrices are added together, forming the full sequence of gate matrices for the embedding layer, which can now be matrix multiplied with the state vector.

The computationally expensive parts of the embedding operation, such as the repeated application of the Kronecker product, are executed only once during circuit initialization and cached for future use. Although the cache is computationally fast, it becomes memory-intensive as the number of qubits increases. To mitigate this, we employ circuit splitting (see \cref{sec:meth:splitting}).

While these matrices consist mostly of zeros (the matrix for $W$ qubits has $2^{2W}-2^W$ zeros), it would be advantageous to use sparse matrix representations, which are faster to multiply with another. However, our preliminary tests have shown that due to the constant multiplications with the quantum state (which is a very dense vector) and the consequent required type conversions, the overhead outweighs the benefits of sparsity. Therefore, Qandle does not utilize sparse matrices for the gate matrices.

PennyLane, on the other hand, employs an aggressive caching approach, where the circuit structure, inputs, and outputs are saved in cache, with structure and input acting as keys. This caching strategy enables fast execution times for repeated executions of the same circuit, particularly when the number of gates and qubits is low. However, as the number of qubits increases, the cache becomes less effective. In many quantum machine learning applications, the input data changes with each forward pass, resulting in frequent cache misses. This further diminishes the benefits of PennyLane's caching mechanism. The impact of PennyLane's caching can be observed in the execution speed comparison presented in \cref{fig:eval:time:mean}.

\subsection{Circuit Splitting}\label{sec:meth:splitting}

One of the major challenges faced by state vector simulators is the exponential growth of the state vector and the corresponding gate matrix size with the number of qubits. As the number of qubits, denoted by $W$, increases, a circuit's state vector size becomes $2^W$, and the gate matrices involved in the computations become $2^W \times 2^W$. Consequently, implementations of quantum circuits that rely on naive state vector and gate matrix multiplications struggle to handle larger circuits efficiently.

To address this computational complexity, we propose a technique called circuit splitting. The idea behind circuit splitting is to divide the circuit into smaller sub-circuits, thereby reducing the matrix sizes and the memory and computation time required. This splitting can be performed during circuit creation, eliminating the need to make a trade-off between splitting quality and execution time. The split circuits, which are essentially groups of quantum gates, can then be executed sequentially, operating only on a subset of the full state vector at a time.

To generate these groups, we interpret the circuit as a dependency graph, where each CNOT gate represents a node, ignoring other gates. In this graph, two CNOT gates are connected by an edge if they share either a control or a target qubit and are successive in the circuit. Currently, our implementation utilizes a simple greedy algorithm. It iterates over all subtrees of the dependency graph and introduces a new group whenever the current group would exceed the given maximum number of qubits (typically between three and six). In the final step, the previously ignored single-qubit gates are added to the nearest group of CNOT gates on the same qubit.

The previously large circuit has been decomposed into smaller sub-circuits, which can be treated as unitary gates acting on multiple qubits. During circuit execution, the state vector is reshaped to match the dimensions of the sub-circuit. After applying the sub-circuit, the state vector is reshaped back to its original dimensions. In the reshaping process, the qubits involved in the sub-circuit are stored in a separate dimension. For example, if the circuit has five qubits labeled $0,1,2,3,4$, and the sub-circuit acts on qubits $1$ and $2$, the reshaping would transform the state vector from $(2^5)$ dimensions to $(d_0\times d_3 \times d_4, d_1 \times d_2)$ dimensions. This allows for matrix multiplication between the sub-circuit (with a gate matrix $G\in\mathcal{C}^{2^2\times2^2}$) and the states over the last dimension. In the case of batched execution, the additional batch dimension of the state vector is merged during reshaping, while storing the original batch size $b$ for reshaping back. This results in a reshaped state vector of dimensions $(b \times d_0 \times d_3 \times d_4, d_1 \times d_2)$ for batched execution. The overhead introduced by this reshaping process has a negligible impact on execution speed compared to the computational load of matrix multiplications. Additionally, hardware caching remains unaffected as the batches are processed independently.

\subsection{Additional Optimizations}\label{sec:meth:opt}

To enhance the quality of the machine learning process, we employ quantum weight remapping techniques \cite{qwmap,qwmap2}. During the remapping process, all quantum weights are transformed to a new range, such as $[-\pi, \pi]$, using smooth functions like the hyperbolic tangent ($\tanh$). The additional computational overhead incurred by the remapping step is negligible compared to the numerous other operations performed during each forward pass. However, it yields noticeable improvements in the training process, including faster convergence and a more stable loss curve \cite{qwmap,qwmap2}.

Furthermore, our investigations suggest that quantum weight remapping can mitigate the negative effects of the barren plateau problem. We provide a detailed analysis of these preliminary findings in \cref{sec:eval:barren}.

In addition, we encourage using PyTorch's \verb|torch.compile| function to further optimize the execution of our simulator.
Since our implementation relies exclusively on PyTorch's tensor operations, it can be compiled into a single execution graph.
This compilation process enables faster execution on both CPU and GPU by optimizing the execution graph. This optimization includes reordering the execution order of parallelizable operations to improve hardware cache layout and fusing consecutive reshaping operations into a single reshaping operation.
By reducing the number of calls to system memory and CPU cycles, the compilation process can significantly enhance the overall performance of our simulator \cite{pytorch}.

\section{Implementation and Evaluation}\label{sec:eval}
\subsection{API}\label{sec:meth:api}
We showcase our proposed techniques by implementing a PyTorch-compatible state-vector simulator. It is designed to ensure compatibility with other quantum software platforms, facilitating easy exporting to the OpenQASM format. In addition, we provide a simple API that closely resembles the standard PyTorch API. This design choice allows for seamless integration of our circuits as \verb|torch.nn.Module|s into existing machine learning workflows. Similar to conventional PyTorch modules such as convolutional layers, we store the quantum weights as parameters, eliminating the need for manual handling of the quantum weights and their gradients, as required in PennyLane. If users still desire to manually access or modify the weights, they can do so using the \verb|parameters| method of the module.

\subsection{Execution Time}\label{sec:eval:time}
The execution time of our proposed methods in our simulator is evaluated by comparing it to the execution times of PennyLane, Qiskit, and TorchQuantum. For PennyLane and Qiskit, which offer multiple backends each, the fastest available backend is chosen for each (determined through pretesting).

To ensure accurate measurements, warm-up runs are performed to allow on-demand/just-in-time compiling of modules, which are then stored in system memory. Random input data is sampled to simulate the execution of a larger dataset which exceed the capacity of CPU caches and system memory. The weights of the quantum circuit are modified using a classical optimizer.
To minimize the influence of other components, a trivial loss function and the well-tested Adam optimizer \cite{adam} are employed.

\begin{figure}[!t]
  \centering
  \includegraphics[width=0.99\linewidth]{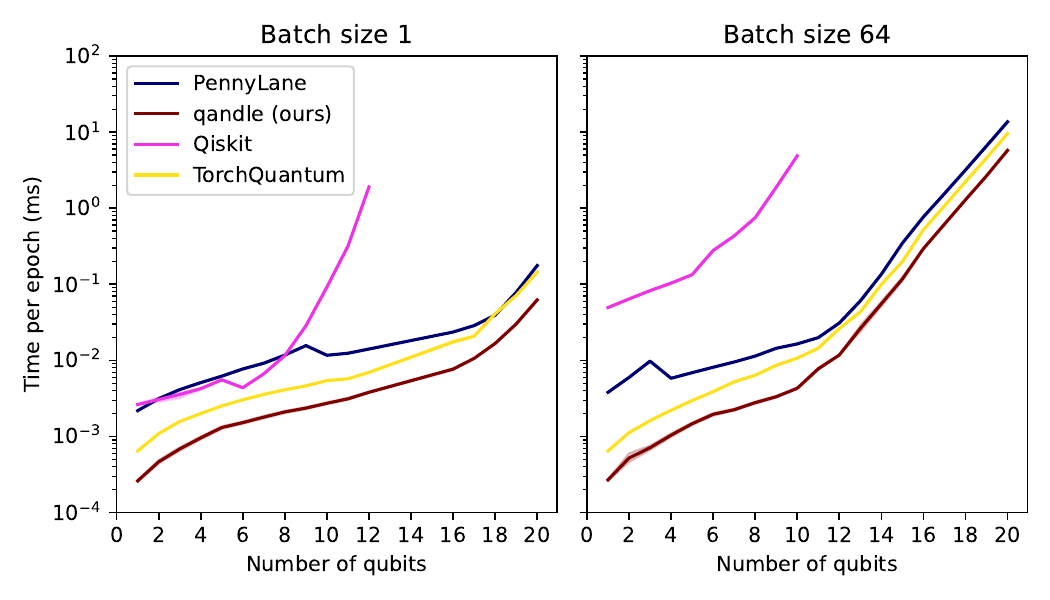}
  \caption{Simulation results for the network.}
  \label{fig:eval:time:mean}
\end{figure}

The evaluation of execution times (mean of 15 runs, other statistics shown in \cref{fig:eval:time:min}) in \cref{fig:eval:time:mean} demonstrates Qandle's superior performance compared to other simulators. Qandle consistently outperforms TorchQuantum, which is specifically designed for high execution speed.

The speed curve reveals the impact of PennyLane's caching mechanism. As the number of qubits increases, the execution times grow until a certain point, determined by the batch size, where the caching feature is disabled. At this point, the execution times briefly decrease before inevitably rising again. This behavior is a result of our experiment setup, which uses different inputs (sampled randomly) and weights (modified by the optimizer) for each forward pass, leading to cache misses. In scenarios where the same circuit is repeatedly executed without changes to the input or weights (e.g., for datasets that fit within the batch size or during inference), PennyLane's caching mechanism would provide better performance than observed in this evaluation. We however argue that this is not a realistic scenario for training a quantum machine learning model.

\subsection{Memory Usage}\label{sec:eval:memory}
To evaluate memory usage, we executed the same circuits on different simulators and measured their peak memory usage. We employed a realistic training scenario, performing multiple backward passes with a simple loss function and varying input data to avoid caching effects. We measured the maximum resident set size (RSS) of the Python process, including the loaded simulator libraries and the PyTorch library, using the GNU \verb|time| command. Each measurement was repeated 15 times, with negligible variance caused by swapping and other system processes. All tests were conducted on workstations with 64 GB of RAM and Intel Core i9-9900 CPUs. Simulators offering multiple backends, such as PennyLane and Qiskit, were executed with their fastest backend variants, \verb|default.qubit.torch| and \verb|statevector_simulator| on the Aer simulator, respectively \cite{pennylane,qiskit}.

\begin{figure}[!t]
  \centering
  \includegraphics[width=0.99\linewidth]{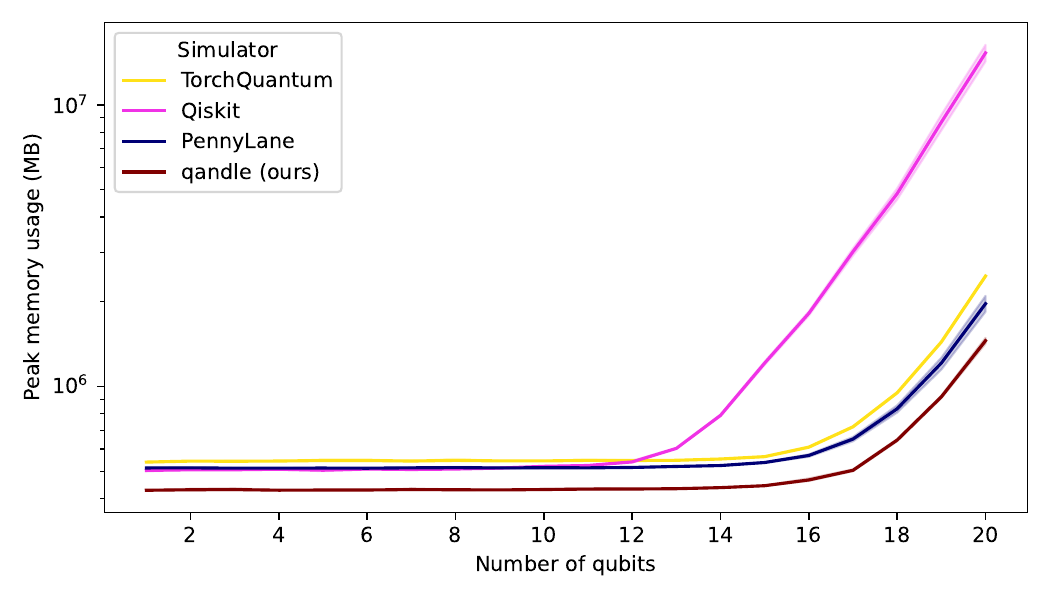}
  \caption{Memory usage for a hardware-efficient SU(2) circuit with varying numbers of qubits. Qandle exhibits lower memory usage compared to other simulators.}
  \label{fig:eval:memory}
\end{figure}

The memory scaling behavior exhibits similar characteristics to other simulators: even with optimizations, memory usage grows exponentially with the number of qubits. This is due to the large size of the state vector, which consists of $2^W$ complex numbers, and the associated memory overhead of matrix multiplications. Over the tested quantum circuits with up to 20 qubits (see \cref{fig:eval:memory} for an implementation of a hardware-efficient SU(2) circuit over all qubits), our simulator demonstrates lower memory usage compared to other simulators, although it still scales exponentially with the number of qubits. TorchQuantum and PennyLane perform similarly (with a slight advantage for TorchQuantum), while Qiskit utilizes the most memory, potentially making it unsuitable for very large circuits.

\section{Conclusion}\label{sec:conclusion}
This paper presents advanced techniques, namely quantum gate matrix caching and circuit splitting, to accelerate the execution of quantum circuits. The showcase implementation, Qandle, is a high-performance state-vector simulator that seamlessly integrates with PyTorch-based machine learning workflows. Qandle demonstrates significant improvements in execution times and memory usage compared to existing frameworks such as PennyLane, Qiskit, and TorchQuantum, validating the effectiveness of the proposed methods. Moreover, Qandle's compatibility with PyTorch's \verb|torch.compile| function further enhances its performance. The user-friendly API of Qandle enables easy integration, even for users with limited experience in quantum machine learning and quantum computing, thereby expanding the accessibility of quantum machine learning to a wider audience.

Based on the promising performance of the proposed methods, we recommend incorporating them into other existing simulators.

As part of future work, we plan to expand the range of supported quantum gates, particularly multi-qubit gates like the Toffoli gate. This expansion will enable the simulation of more complex circuits that are currently not supported by our implementation. Additionally, we aim to develop a more sophisticated splitting algorithm based on graph algorithms, leveraging the circuit's dependency graph. This algorithm will determine the optimal split, reducing the number of sub-circuits and minimizing the overhead of reshaping the state vector, while ensuring efficient execution. We propose exploring graph coloring techniques or split decomposition algorithms for this purpose.

\appendices

\section{Expanded Evaluation}

\subsection{Barren Plateaus}\label{sec:eval:barren}
\begin{figure}[!t]
  \centering
  \includegraphics[width=0.99\linewidth]{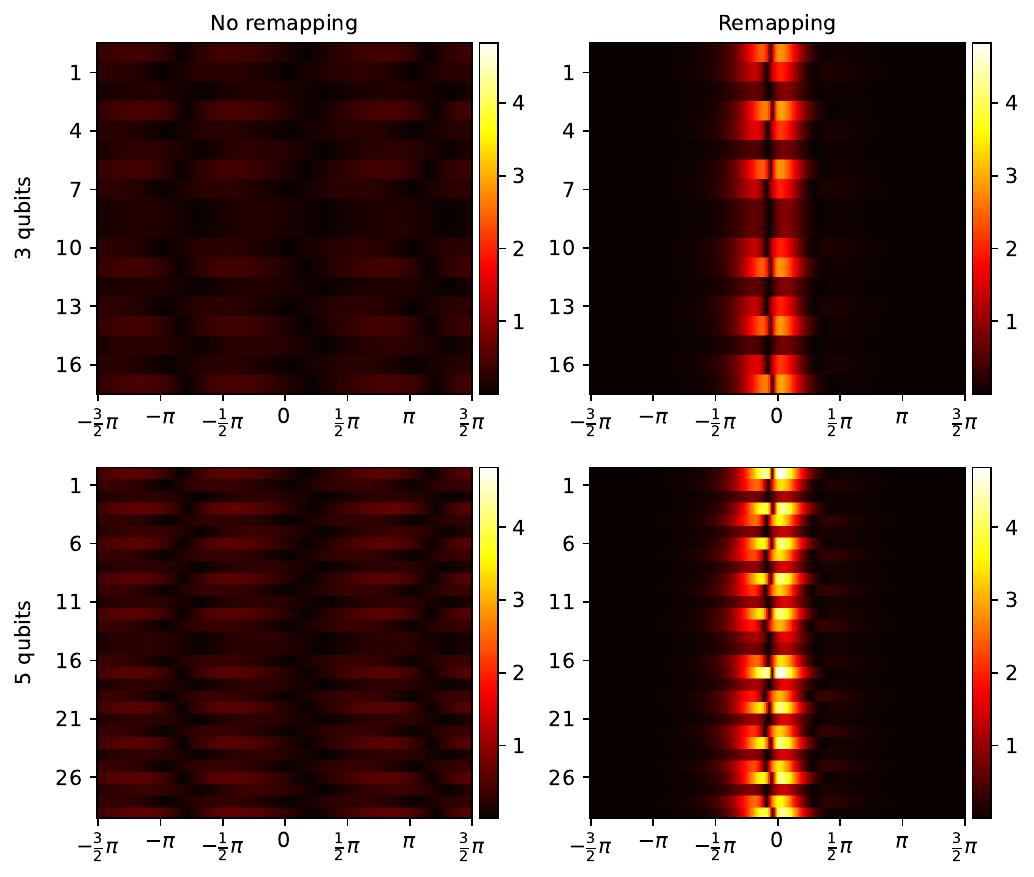}
  \caption{Gradients in circuits with 3 and 5 qubits. The y-axis represents the parameter index, while the x-axis displays the absolute gradient of that parameter at different rotations, with other parameters remaining constant. Brighter colors signify higher absolute gradients.}
  \label{fig:eval:barren}
\end{figure}

The phenomenon of barren plateaus, where the gradients of the loss function become exponentially small in relation to the number of qubits and the depth of the circuit, is a well-documented challenge in quantum machine learning \cite{barren,barreninit}. As illustrated in \cref{fig:eval:barren}, we posit that the impact of barren plateaus can be mitigated through the application of quantum weight remapping, which effectively constrains the weights within a narrower range. This hypothesis was tested on Strongly Entangling Layers (which have been proposed by \cite{sel}), with varying numbers of qubits. Contrary to the method proposed by \cite{barreninit}, which suggests initializing the weights using the inverse of their predecessors for random circuits, we found that random initialization of weights is sufficient, given that remapping will subsequently constrain them within a smaller range. In \cref{fig:eval:barren}, areas of higher gradients are represented by brighter colors. Dark spots surrounded by bright colors signify local optima, while uniformly dark colors indicate barren plateaus. Due to the restriction of the weights, the gradients are more pronounced, and barren plateaus are absent between the local optima. The remapped solution landscape exhibits smoothness within the range of $\left[-\frac{1}{2}\pi,\frac{1}{2}\pi\right]$. Interestingly, a higher number of qubits results in even greater gradient amplitudes when combined with quantum weight remapping. We recommend further research to explore the impact of quantum weight remapping on the barren plateau problem in greater depth, including the examination of random circuits and different initialization methods. 

\subsection{Execution Time}
\begin{figure}[!t]
  \centering
  \includegraphics[width=0.99\linewidth]{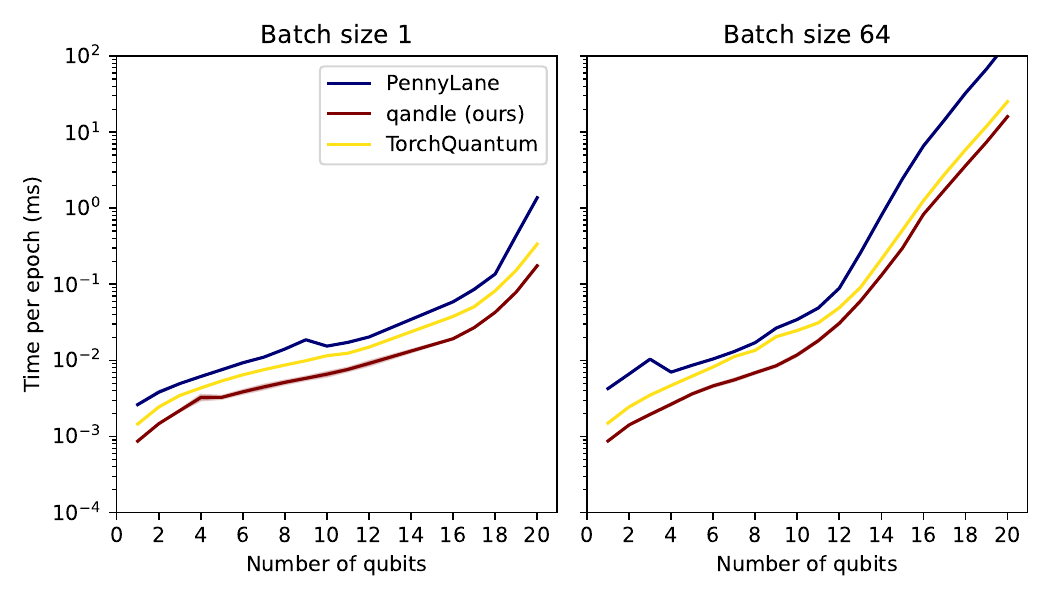}
  \caption{Simulation results for the network, showing only the fastest run.}
  \label{fig:eval:time:min}
\end{figure}

In \cref{fig:eval:time:min}, we show the minimum execution times of the simulators for the same circuits, employing a full forward and backward pass. The results are consistent with the mean execution times shown in \cref{fig:eval:time:mean}, showing Qandle as the fastest simulator, followed by TorchQuantum and PennyLane. Minimal execution times are more effected by other system processes and caching mechanisms, and are therefore less reliable to reproduce.

\ifCLASSOPTIONcaptionsoff
  \newpage
\fi

\bibliographystyle{IEEEtran}
\bibliography{bibliography}

\end{document}